\def\apj{{ApJ}}

\def\mnras{{MNRAS}}

\documentclass[cup5b]{caps}
\usepackage{graphicx}
\usepackage{amssymb}
\usepackage{ociwsymp1e}
\HeadText{M. Lacy}

\begin{document}

\pagenumbering{arabic}

\author[]{M. LACY\\SIRTF Science Center, Caltech}

\chapter{Massive Black Holes at High \\ Redshifts}

\begin{abstract}
Black holes with masses $M_{\rm bh} \sim 10^{8.5} M_{\odot}$ dominate
the accretion history of the Universe. These black hole masses are typical
of those found in radio-selected galaxies today, suggesting that the giant 
elliptical hosts of low redshift radio galaxies were the hosts of 
powerful, mostly radio-quiet, quasars in the high redshift Universe. The 
reason that all radio 
galaxies are found in such hosts may be the correlation of black hole mass
with radio luminosity, but it is emphasized that accretion rate too plays 
an important role in the production 
of powerful radio jets. The tight $K-z$ relation of luminous, high redshift, 
radio galaxies is probably a selection effect due to the selection on 
high black hole masses and high accretion rates. Luminous radio galaxies are
the radio-loud part of the ``quasar-2'' population, and the ratio of 
radio-loud quasars to luminous radio galaxies, about 1:1, is so far our
only good estimate of the relative numbers of quasar-1s and quasar-2s. The 
numbers of radio-quiet and radio-intermediate
quasar-2s are still uncertain, but a much larger 
population than the quasar-1s would conflict with 
constraints from the present-day
black hole mass density. A comparison of the number densities of dark matter
haloes and the high redshift quasars, however, suggests that 
there are plenty of dark haloes capable of hosting 
the known high redshift AGN and thus room for a significant, but not 
huge, quasar-2 population.
\end{abstract}

\section{Introduction}

The most massive black holes ($\stackrel{>}{_{\sim}} 10^8 M_{\odot}$) are of
particular interest as it is the quasars containing these black holes which 
contribute most to the integrated accretion luminosity of the Universe at 
the height of the ``quasar epoch'' at $z\sim 2$. Their hosts today, 
giant ellipticals, are also typically seen as
the hosts of nearby luminous quasars and radio galaxies. 
Furthermore, even out to $z\sim 4$ radio galaxies are still 
the most luminous galaxies known at any redshift. In this paper I begin by 
discussing correlations between radio luminosity and the properties of 
the AGN (black hole mass, accretion rate and (possibly) spin). I then 
discuss the radio galaxy $K-z$ relation and its implications for the nature
of radio-loud AGN and their host galaxies. Another point about which there 
was some discussion at the meeting was the obscured ``quasar-2'' 
population, so I have 
added a brief survey of the current state of radio, IR and X-ray 
selection of dust reddened/obscured quasars. 
Finally, I conclude with a discussion of how we
might relate the number densities of high mass black holes as measured 
by AGN luminosity functions to the number 
densities of massive dark matter haloes at high redshift, and 
briefly discuss the uncertainties in this calculation.

\section{The Production of Powerful Radio Jets}

\subsection{Black hole mass}

The ability to estimate black hole masses of AGN from H$\beta$ linewidths
has provoked a large 
number of papers which attempt to relate black hole mass 
to the radio properties of the AGN, with varying degrees of success
(Laor 2000; Lacy et al.\ 2001; Boroson 2002; Ho 2002; 
Oshlack, Webster \& Whiting 2002; Woo \& Urry 2002). The relationship
clearly has a lot of scatter; at least at low redshift, all luminous AGN 
are found in massive galaxies with correspondingly massive black holes, and
the difference in the mean black hole masses between radio-loud and 
radio-quiet
objects (if any) is small. Weak intrinsic correlations can easily 
be masked through 
orientation effects, however, due to a 
combination of both beamed radio emission and 
orientation-dependence of broad-line widths (Jarvis \& McLure 2002). 
Nevertheless, it is also 
true that samples in which correlations have been found, although consistently
selected, complete, and mostly consisting of steep-spectrum objects free
from orientation bias, have been fairly
small ($<100$ quasars). Deeper radio surveys, combined with the Sloan quasar
survey, are probably needed before a definitive statement can be made on the
basis of linewidths alone. However, host galaxies can provide good, 
orientation-independent estimates of black hole masses, either from the 
host magnitudes or (in the case of nearby radio galaxies and BLLacs) 
from measurement of the velocity dispersion of the host galaxy,
and all the evidence so far 
from such studies points to radio-luminous AGN being found only in the most
massive galaxies, with massive black holes (e.g.\ McLure et al.\ 1999; O'Dowd,
Urry \& Scarpa 2002; Barth, Ho \& Sargent 2003; Bettoni et al.\ 2003). 
For example, FRI radio sources, with very low accretion 
powers, but relatively high radio luminosities (and very high radio-loudnesses) are always hosted by giant elliptical galaxies, whereas Seyfert galaxies 
and LINERs with similar accretion luminosities exist in a large range of 
hosts, on average less massive. 

\subsection{Accretion rate}

The role of accretion rate is well-illustrated by the 
radio luminosity -- emission line luminosity correlation for
steep-spectrum radio galaxies, which extends over five orders of magnitude
in radio luminosity (e.g.\ Willott et al.\ 1999; Figure 1.1). 
With only one or two exceptions in samples totalling several hundred, all 
such radio galaxies are hosted by giant ellipticals with luminosities of 
$L^{*}$ and above, and thus have a very narrow range of black 
hole masses ($10^{8}-10^{9} M_{\odot}$). The range in radio luminosity must
therefore derive principally 
from the range in accretion rate. Interestingly, the 
correlation is only close to linear at the highest radio/emission line
luminosities, and at low luminosities the logarithmic slope of the 
correlation decreases to about 0.3 (Zirbel \& Baum 1995). This suggests that 
radio jet production could become more efficient at low accretion rates,
leading to the anticorrelation of radio-{\em loudness} (i.e.\ the ratio 
of radio luminosity to AGN accretion luminosity) and accretion rate 
relative to the Eddington limit noted by Ho (2002). 
Samples of local radio galaxies, with their small 
range of black hole masses
but large range of accretion rates show no correlation of black hole mass
with radio luminosity (Bettoni et al.\ 2003), as expected if their range in 
radio luminosities is dominated by their range of accretion rates.

  \begin{figure}
    \centering
\includegraphics[scale=0.7]{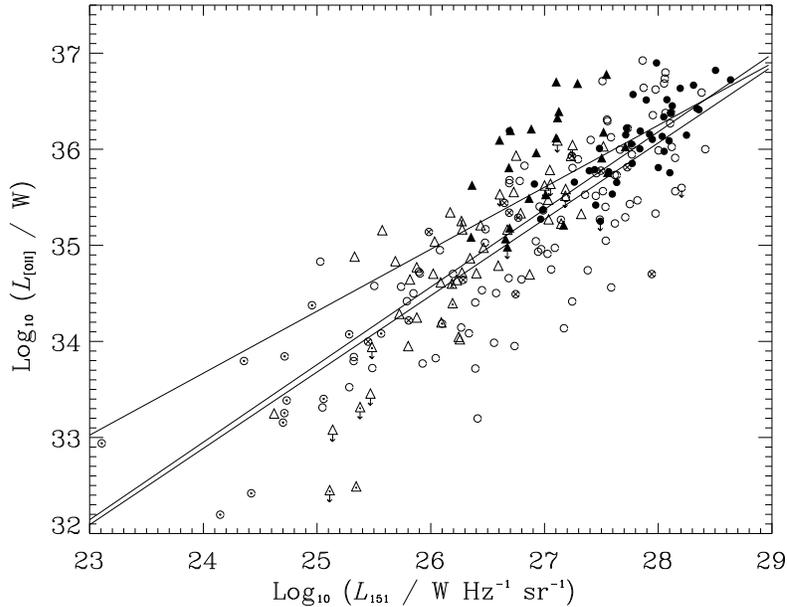}
    \caption{The radio luminosity -- [OII] luminosity plot for radio galaxies
and quasars 
from Willott et al.\ (1999). 3C radio sources are represented by circles and
7C sources (selected at a radio flux level $\approx 20$ times lower than 3C) 
are shown as triangles. Open symbols represent radio galaxies, open symbols
with crosses weak quasars (i.e.\ mostly broad-line radio galaxies), open 
symbols with dots FRI radio galaxies and solid symbols quasars.
The two steeper lines are fits to the high 
luminosity radio sources only (one for radio galaxies and one for both radio
galaxies and quasars), the line with the flatter slope is a fit to 
all the FRII radio galaxies.}
    \label{sample-figure}
  \end{figure}

\section{The $K-z$ Relation for Radio Galaxies}

Ever since the early 1980's when Lilly \& Longair (1982) first published their
$K$-band photometry of powerful 3C radio galaxies, the cause of the good
correlation between $K$-band flux and redshift, and its implication -- that 
radio galaxies exist only in the most massive galaxies -- has
been a puzzle. This is not only true locally, but continues to be 
the case out to at least $z\sim 4$
(de Breuck et al.\ 2002; Jarvis et al.\ 2002).
Initially, it was thought that emission-line and nuclear contamination by the 
AGN might dominate the $K$-band emission of high-$z$ radio galaxies, but, 
although this seems to be true in a few cases (e.g.\ Eales \& Rawlings 1993),
it does not seem to be the case in general. 
The advent of redshift surveys of radio source samples significantly
fainter than 3C, where AGN contamination of the $K$-band light is negligible,
allowed us to show that the bulk of the light in high-$z$ radio galaxies
is indeed stellar (Lacy, Ridgway \& Bunker 2000; Willott et al.\ 2002), 
and this has recently been 
confirmed by HST NICMOS imaging of the 3C sample itself, where the resolution
is high enough to separate AGN-related emission from the stellar light
(Zirm et al.\ 2003). 
As has been pointed out before (Rawlings 2003; Kukula et al.\ 2001), this 
is consistent with the observed correlation of black hole mass and radio 
luminosity (most radio galaxies are steep spectrum objects, and thus avoid
the complications of  beaming). With the quantitative estimate of the 
dependence of radio luminosity on black hole mass from Lacy et al.\ (2001),
we can predict the distribution of black hole masses in a sample of 
radio-selected objects with radio luminosities above a given threshold. Figure
1.2 shows this calculation for 5GHz radio luminosity cuts of $10^{24,25,26,27} 
{\rm WHz^{-1}sr^{-1}}$, compared to the black hole mass distribution
obtained from the velocity dispersion data for 3C radio galaxies of 
Smith et al.\ (1990). Pushing to the 
higher radio luminosities of high redshift radio galaxies narrows the 
distribution further, as the Schechter function cutoff in the black hole mass
distribution comes into play, and the accretion rates approach the Eddington 
limit, explaining the tightness of the $K-z$ relation. This also explains
the radio-luminosity dependence of the host galaxy magnitudes. Selection 
effects mean that all high-$z$ radio galaxies have similar, high, accretion 
rates relative to the Eddington limit, leaving black hole mass as the only
variable.

  \begin{figure}
    \centering
\includegraphics[scale=0.4]{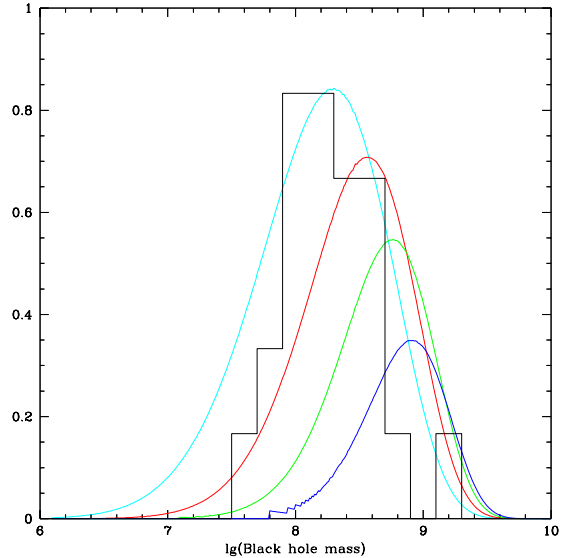}
    \caption{The curves show the predicted distribution of black hole masses
for radio galaxy samples selected above radio luminosity limits of 
$L_{5GHz} =10^{24} {\rm WHz^{-1}sr^{-1}}$ (cyan), $L_{5GHz} =10^{25} 
{\rm WHz^{-1}sr^{-1}}$ (red), $L_{5GHz} =10^{26} {\rm WHz^{-1}sr^{-1}}$ 
(green), and $L_{5GHz} =10^{27} {\rm WHz^{-1}sr^{-1}}$ (blue). The solid
line is the distribution of radio galaxy black hole masses in the Smith
et al.\ (1990) sample with a luminosity 
limit of $L_{5GHz} =10^{24} {\rm WHz^{-1}sr^{-1}}$, which can be
directly compared to the cyan curve.}
    \label{sample-figure}
  \end{figure}

Differences between radio-loud and radio-quiet
AGN hosts seem particularly pronounced at high redshift,
where both Kukula et al.\ (2001) and Ridgway et al.\ (2001)
note a 2 magnitude difference between the host galaxies of
radio-loud quasars/radio galaxies and radio-quiet quasars. Why this should 
be so much stronger an effect at high redshift than locally is unclear.

\section{Dust-reddened and Obscured Quasars}

One of the fundamental tests of the ``standard model'' -- 
that black holes form 
by unobscured radiative accretion of matter -- is that the total mass density 
in black holes today, times a plausible accretion efficiency, should equal 
the integrated luminosity density from quasars. Recently, Yu \& Tremaine 
(and Yu, these proceedings) have updated the original  
analysis of Soltan (1982) using the early-type galaxy luminosity 
function from the Sloan Digital Sky Survey 
(SDSS). Their analysis suggests that the agreement between these two numbers
is remarkably good. This good agreement would suggest that there is little
room for a substantial population of quasars whose apparent luminosity
is decreased by dust obscuration, apparently consistent with the small 
numbers of reddened quasars found, even in the SDSS. Dust-reddened quasars 
are, however, notoriously difficult to find, being strongly selected against 
in any optically-based, magnitude-limited survey (e.g.\ Lacy et al.\ 2002),
and it is possible that many more intrinsically luminous quasars 
remain to be discovered. Certainly, in the parts of parameter space where
we have been able to define ``complete'' populations of AGN independent of 
reddening towards the nucleus, namely, IRAS-selected Seyfert galaxies and 
radio-luminous AGN, we typically find significant populations
of AGN whose quasar nucleus is obscured by absorbing columns of dust, with
rest-frame extinctions in the range $A_V\sim 1-30$. (For simplicity I'll refer
to the whole of this population as ``quasar-2s'' even though some of these
objects with lower extinctions often have broad lines visible in 
near-infrared or deep optical spectra.)

\subsection{Current estimates of the numbers of quasars-2s}

\subsubsection{Radio surveys}

Radio-loud quasar-2s have been known for many years in the form of
radio galaxies. Although it now seems clear that, at lower luminosities,
not all radio galaxies contain hidden quasars, most radio galaxies 
with radio luminosities above 
$L_{\rm 5GHz}\sim 10^{25.5} {\rm WHz^{-1}sr^{-1}}$ 
contain hidden quasars, and comprise about
half the total radio source population at this radio luminosity 
and higher (Willott et al.\ 2000). 
Radio surveys are unbiased with respect to dust reddening, but 
even relatively deep surveys such as FIRST do not go 
deep enough to find the bulk of the quasar population. 
By matching the 2MASS near-infrared survey with the FIRST radio survey, and
searching for objects which disappear on the Palomar Observatory Sky Survey 
we have been successful at finding radio-intermediate quasars with moderate 
rest-frame reddenings (e.g.\ Lacy et al.\ 2002; Gregg et al.\ 2001). 
Most quasars-2s will be missing 
even in 2MASS (our study is most sensitive to $1<A_V<3$ for quasars with 
$0.5<z<3$, compared to a typical Seyfert 2 or radio galaxy reddening of 
$A_V \sim 10$ towards the AGN), but comparing the fractions of 
moderately-obscured quasars in the
3CRR radio-loud sample (Simpson, Rawlings \& Lacy 1999) 
with our 2MASS findings indicates that the fraction of obscured
quasars would be similar if the distributions of rest-frame reddenings are 
the same in our FIRST-2MASS sample as they are in 3CRR. This would imply that 
about half of all radio-intermediate quasars are dust-reddened.

\subsubsection{Hard X-ray surveys}

Observations with Chandra, in particular the deep surveys, have revealed
large numbers of AGN, but typically these have been low luminosity objects,
with a few notable exceptions (e.g.\ Norman et al.\ 2002).
One of the reasons for this may be simply that the Chandra deep surveys cover 
very little sky area, and thus sample relatively small volumes and 
expect to find very few luminous objects. Crawford, Ghandi
\& Fabian (2003) use a serendiptous Chandra survey to cover a much 
larger sky area and can thus find rare objects like high luminosity AGN. This
survey has already found several plausible quasar-2s. As Fabian
(these proceedings) and Gandhi \& Fabian (2003) discuss, 
however, the evidence from the deep surveys
suggests that the obscured AGN are predominately of low luminosity, and 
thus accretion onto them contributes little to the local mass density of 
black holes. 
We will probably have to wait until Chandra serendipitous 
surveys find a statistically-useful sample of obscured quasars before this 
can be accurately determined.

\subsubsection{IR surveys}

Cutri (2003) and collaborators
have used a near-IR color criterion to successfully select 
dust-reddened quasars from the 2MASS survey, but their technique is 
again limited to a fairly narrow range in rest-frame reddening. 
Fortunately, the hot dust component in obscured quasars 
will be very visible in the mid-IR. 
Already, a red FeLoBAL quasar has be found by ISO (Duc et al.\ 2002) 
and many more should appear in the SIRTF surveys. Although the most 
obscured quasars in the SIRTF surveys will be hard to differentiate from
starbursts on the basis of IR/optical SEDs alone, moderately obscured
quasars with $A_V\sim 1-5$ should stand out on the basis of near to mid-IR 
excesses from hot dust emission near the nucleus.

\subsection{How many quasar-2s might there be?}

If the fraction of obscured radio-quiet quasars is similar to the fraction
in radio-loud samples, 
then the additional factor of two in quasar luminosity density
would probably not be a major
problem for the Soltan argument, given the uncertainty in, e.g., relating 
bulge luminosity to black hole mass. There is one good reason to expect more
radio-quiet quasars to be obscured, however. There are more radio-quiet quasars
on the high-accretion rate end of Eigenvector 1 (Boroson 2002), and high 
accretion rates are typically linked to absorption by Broad Absorption Line
(BAL) systems, and thus to excess reddening (Sprayberry \& Foltz 1992). 

\section{Number Densities of High-Redshift Black Holes}

One potentially exciting consequence of the relationship between 
galaxy velocity dispersion and black hole mass is the ability to directly
tie AGN populations to the number densities of the massive black holes 
hosting them. Since the pioneering work of Efstathiou \& Rees (1988), 
several calculations have used progressively more refined 
versions of the Press-Schechter method to derive dark halo mass and 
temperature/velocity distributions as a function of redshift. Recently 
Mo \& White (2002) have computed these for a $\Lambda$-dominated cosmology.
Most relevant to our work is the halo velocity distribution, $\sigma_{DM}$,
rather than the halo mass, $M_{DM}$. As Mo \& White point out, 
$M_{DM}$ decreases with 
increasing redshift $\propto (1+z)^{-1.5}$ even in the absence of halo mergers
and accretion, because the 
cutoff radius within which the halo mass is measured is defined as 
a fixed overdensity relative to 
the cosmic mean at redshift $z$, and thus increases as $z$ decreases. 

To compare the numbers of dark haloes and black holes, however,
we need to know how the black hole mass -- velocity
dispersion relation evolves, and how to relate the host galaxy velocity
dispersion ($\sigma_{s}$) to $\sigma_{DM}$. Progress
on the first of these problems is being made fairly rapidly. The width of 
the [OIII]5007 emission line can be used as a crude measure of $\sigma_{s}$, 
though it is clear that there is 
a lot of scatter in this relation (Shields et al.\ 2003; Boroson 2003; 
Heckman, these proceedings). Also promising is the use of quasar host 
luminosities to 
estimate black hole masses independent of AGN-related emission lines using 
HST or AO data (Dunlop, 
these proceedings, Lacy et al.\ 2002b). Progress on the second question,
the relationship of $\sigma_{s}$ to $\sigma_{DM}$, is 
more difficult. This is because the dark matter in the cores of the dark
matter haloes needs to be replaced with the baryonic matter that dominates
the mass distribution in the centers of early-type galaxies today via a 
mechanism that is not yet understood.
Kochanek (1994) examined this problem observationally by fitting the velocity
dispersion profiles of 37 nearby elliptical galaxies, and found no 
significant differences between $\sigma_{s}$ and $\sigma_{DM}$, but his 
modeling assumed only a simple isothermal 
sphere model for the dark matter. Ferrarese (2002) and Wyithie \& Loeb
(2002) examined this problem from a theoretical point of view, and obtained
essentially the same result for $L^{*}$ galaxies, namely that 
$\sigma_{s} \approx \sigma_{DM}$, though predicting a 
non-linear relationship which diverges from this for more or less massive 
galaxies. Somerville (these proceedings), however, suggests 
that current hierarchical galaxy formation models predict no 
straightforward relation between the two velocity dispersions, and that one 
might expect a large scatter depending on the exact process by which dark 
matter in the halo core is replaced by the baryonic matter which dominates the 
inner parts of galaxies today.

  \begin{figure}
    \centering
\includegraphics[scale=0.5]{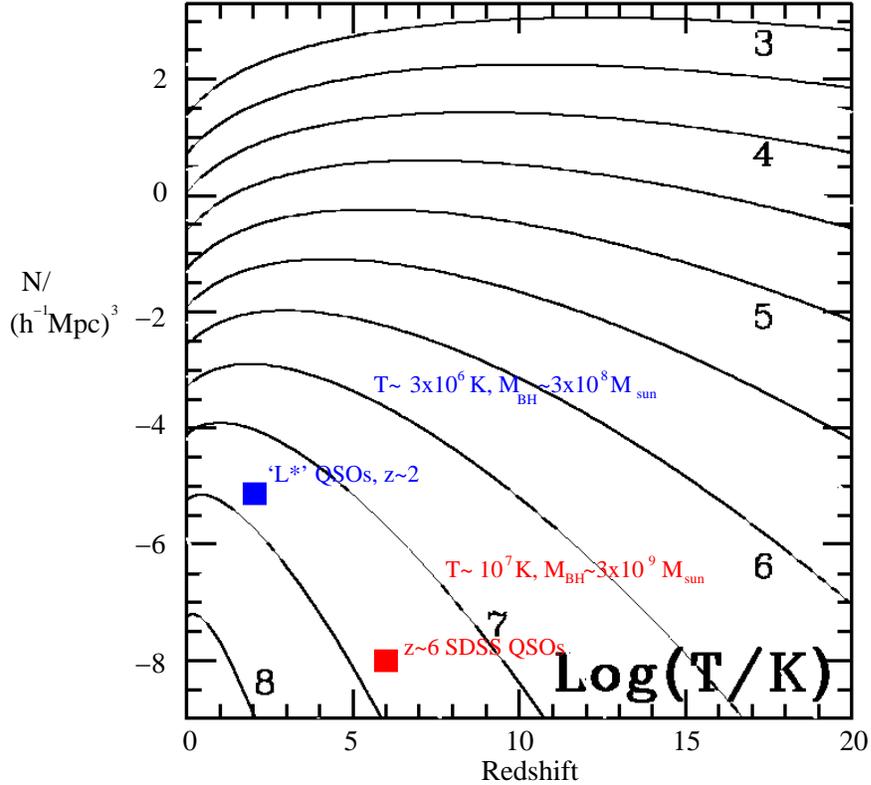}
    \caption{The upper right panel of figure 2 from Mo \& White (2002)
with annotations. The solid lines show the number densities of dark 
matter haloes as a function of redshift for halo temperatures of 
${\rm lg}(T/K)= 3-8$ (labeled). The blue square is the approximate
number density of $L^{*}$ quasars at $z\sim 2$, and the red square the 
approximate number density of $z\sim 6$ quasars from the SDSS. Corresponding
dark halo number density curves are indicated by the blue and red labels
for the $z\sim 2$ and the $z\sim 6$ quasars respectively, assuming no 
evolution in the $M_{bh} - \sigma$ relation and $\sigma_{DM} \approx 
\sigma_{s}$ (see text for details).}
    \label{sample-figure}
  \end{figure}

If we assume for the moment that dark matter and stellar velocity dispersions
are similar, we can use the current best estimates of the quasar number 
density to compare with the number of dark haloes, and, assuming 
accretion at the Eddington rate, we can thus obtain an estimate of the 
duty cycle and typical quasar lifetime to compare with the independent
estimate obtained from the Soltan argument by Yu \& Tremaine (2002). 
It turns out (Figure 1.3) that there are $\sim 100$ times as many dark haloes
capable of hosting $L^{*}$ quasars as there are $L^{*}$ quasars at $z\sim 2$, 
implying a duty cycle of $\sim 0.01$ during the $\sim 10^9$yr 
``quasar epoch'' and hence a
quasar lifetime $\sim 10^{7}$yr. This compares with the estimates of Yu \& 
Tremaine (2002), who estimate lifetimes for $L^{*}$ quasars $\sim 3\times 
10^{7}$yr. This relatively small 
difference may be due to stellar velocity dispersion indeed being a 
poor estimator of dark matter velocity dispersion, or the local estimate
of black hole mass density being low, 
e.g.\ because black holes are occasionally
expelled from galaxies during 3-body interactions, or simply because the 
$M_{bh} - \sigma$ relation is still a little in error. Taken at face 
value, the halo numbers do leave some
room for a quasar-2 population as much as a few times larger than the quasar-1
population, but a quasar-2 population an order of magnitude or larger than
the quasar-1 population can probably be ruled out both on the basis of the
Soltan argument and the numbers of available dark haloes. 

Another interesting feature of Figure 1.3 comes from comparing the number
densities of haloes capable of hosting highly-luminous quasars at $z>6$ and 
the number density of such quasars found in the SDSS. There are $\sim 100$
times as many haloes as observed quasars. As these
black holes must be accreting continuously close to the Eddington rate to 
form their very high mass black holes by $z\sim 6$, one 
might speculate that a large population of obscured quasars
with very massive black holes could exist at high redshifts.

The number densities of radio galaxies and radio-loud quasars at $z\sim 2$ 
are comparable to the number densities of optically-selected quasars of 
similar black hole mass (as estimated from their host galaxy luminosities). 
The radio galaxies typically have much 
lower accretion rates, however, suggesting that they are the less active, but 
perhaps longer-lived, radio-bright equivalents of the quasars. 

\section{Conclusions}

The ability to estimate black hole masses in AGN has provided us with an 
excellent opportunity to investigate correlations between AGN properties such
as the presence of powerful radio jets or BAL flows with black hole mass
and accretion rate. Although such studies are in their early phases and 
sparking much controversy, it seems likely that some answers to basic questions
(Why are radio galaxies always giant ellipiticals whereas radio-quiet AGN 
can exist in spirals? Why is the $K-z$ relation for radio galaxies so tight?
Is there a link between BAL flows and high accretion rates? Why are radio-loud 
BAL quasars rare?) can come from such studies using samples selected to 
cover appropriate ranges in parameter space. A speculative ``Grand Unified
Scheme'' for AGN using black hole mass, accretion rate and orientation
as the principal variables can even be constructed (Lacy 2003). 

The tight constraints on integrated quasar number densities at the peak of 
the quasar epoch place strong constraints on the allowed fraction of 
quasar-2s. Nevertheless, such objects continue to be found with the new 
generations of all-sky, large area, multi-wavelength surveys, and what 
little evidence we have so far suggests that true quasar number densities 
could be at least a factor of two higher than those of quasars detected
in optical and soft X-ray surveys.

The use of high redshift AGN for constraining galaxy formation models has 
also received a boost from the ability to estimate black hole masses from 
broad line widths  (e.g.\ Vestergaard, these proceedings). 
Relating the dark matter velocity dispersion to that of the galaxy which forms
inside it remains a problem which needs to be further addressed both 
observationally and in galaxy formation models. However, it seems clear that
very massive black holes existed early in the history of the Universe, making 
galaxy formation models which form large galaxies early (e.g.\ 
Granato et al.\ 2001; Loeb \& Peebles 2003) particularly attractive.

\section{Acknowledgements} 
This work was carried out at the Jet Propulsion Laboratory, California 
Institute of Technology, under contract with NASA.

%





\end{document}